# The human visual system and CNNs can both support robust online translation tolerance following extreme displacements


Ryan Blything[1*], Valerio Biscione[1], Ivan I. Vankov[2], Casimir J.H. Ludwig[1],

Jeffrey S. Bowers[1]

[1]School of Psychological Science, University of Bristol, Bristol, United Kingdom

[2]Department of Cognitive Science and Psychology, Sofia, New Bulgarian University, Bulgaria

**\*Corresponding author:** Email: j.bowers@bristol.ac.uk



**Author contributions:** Conceptualization and Writing: RB, VB, IV, CL, JB. Investigation: RB (psychophysical experiments) and VB, IV (CNN simulations).

**Declarations of interest:** none

**Funding:** This project has received funding from the European Research Council (ERC) under the European Union's Horizon 2020 research and innovation programme (grant agreement No 741134).






**Abstract**

Visual translation tolerance refers to our capacity to recognize objects over a wide range of different retinal locations.  Although translation is perhaps the simplest spatial transform that the visual system needs to cope with, the extent to which the human visual system can identify objects at previously unseen locations is unclear, with some studies reporting near complete invariance over 10° and other reporting zero invariance at 4° of visual angle. Similarly, there is confusion regarding the extent of translation tolerance in computational models of vision, as well as the degree of match between human and model performance. Here we report a series of eye-tracking studies (total N=70) demonstrating that novel objects trained at one retinal location can be recognized at high accuracy rates following translations up to 18°.  We also show that standard deep convolutional networks (DCNNs) support our findings when pretrained to classify another set of stimuli across a range of locations, or when a Global Average Pooling (GAP) layer is added to produce larger receptive fields. Our findings provide a strong constraint for theories of human vision and help explain inconsistent findings previously reported with CNNs.

**Keywords**

Translation Tolerance; Translation Invariance; Object Recognition; Convolutional Neural Networks; Global Average Pooling





## 1   Introduction

We can identify familiar objects despite the variable images they project on our retina, including variation in image size, orientation, illumination, and position. How the visual system succeeds under these conditions is still poorly understood. Here we consider the case of variation across position and the extent to which the visual system and artificial neural networks can identify objects at previously unseen locations.

Although translation is perhaps the simplest spatial transform that the visual system needs to cope with, there is nevertheless confusion in the literature regarding the extent of translation tolerance in human vision, the extent of tolerance in computational models of vision, as well as the degree of match between humans and models.  There are both theoretical and methodological reasons for this confusion.  With regards to theory, researchers often fail to distinguish between online tolerance and trained tolerance (see Bowers, Vankov, & Ludwig, 2016). In the case of *online* tolerance, learning to identify an object at one location *immediately* affords the capacity to identify that object at multiple other retinal locations even when no members of that category have ever been seen at those other locations. For example, if a person has only seen dogs projected at one retinal location they may nevertheless be able to identify dogs when projected to other retinal locations.  At one extreme, the visual system can immediately generalize to *all* locations (within the limits of visual acuity), what might be called on-line translation invariance; at the other extreme, there is no generalization to untrained locations.  *Trained* tolerance, by contrast, refers to the hypothesis that we can identify a novel exemplar of an object category across a range of retinal locations by training the visual system to identify other exemplars of this category





across a broad range of retinal locations.  For instance, learning to identify multiple images of dogs at multiple retinal locations allows the visual system to identify a new image of a dog across multiple locations.  This distinction is often ignored in discussion of translation tolerance in humans and computational models of vision, leading to a wide range of different conclusions regarding the extent of translation tolerance.  Here we are concerned with online translation tolerance.

Furthermore, when behavioural studies were specifically designed to assess on-line translation tolerance, a variety of methodological differences has led to a wide range of findings, ranging from no on-line tolerance at 4° (Cox & DiCarlo, 2008) to near complete on-line tolerance at 13° (Bowers et al., 2016).  Similarly, different computational models of object classification support varying degrees of online translation tolerance, from near zero tolerance (Chen et al., 2017) to complete invariance (Han et al., 2020).  These mixed outcomes have led to contrasting conclusions, with many researchers emphasizing the importance of trained rather than on-line tolerance in both biological and computational models of vision (e.g., Cox and DiCarlo 2008; Dandurand et al. 2013; Di Bono and Zorzi 2013; Edelman and Intrator 2003; Elliffe et al. 2002; Serre, 2019), and others proposing theories that support on-line translation invariance (Biederman, 1987; Hummel & Biederman, 1992).  Researchers have also modified the architectures of standard convolutional neural networks (CNNs) in order to explain the lack of translation tolerance beyond 4° (Chen et al. 2017), or alternatively, modified CNNs in order to explain the near complete translation tolerance in some conditions and limited translation tolerance in others  (Han et al., 2020).  In the current work we show that on-line translation tolerance for images of novel 3D objects is greater in human vision than previously demonstrated even when the images are flashed for





100 ms and masked at an untrained retinal position at test. In addition, our simulations highlight several conditions in which CNNs demonstrate extreme translation tolerance and display human-level like performance.

## 1.1 Brief review of on-line translation tolerance in biological and artificial visual systems

In the case of biological vision, early behavioural studies provided evidence for robust translation tolerance following 10° of translation based on long-term priming studies (Biederman & Cooper, 1991; Cooper, Biederman, & Hummel, 1992; Ellis, Allport, Humphreys, & Collis, 1989; Fiser & Biederman, 2001; Stankiewicz & Hummel, 2002; Stankiewicz, Hummel, & Cooper, 1998). Similarly, in single-cell neurophysiological studies, researchers have identified neurons in inferior-temporal cortex (IT) with extremely large receptive fields (up to 26°; Op De Beeck, & Vogels, 2000) that are thought to provide the neural underpinning of translation tolerance. However, these findings were obtained with familiar stimuli, and this has led researchers to argue that robust priming and large receptive fields might reflect trained rather than on-line translation tolerance (e.g., Kravitz, Vinson, & Baker, 2008). That is, although the specific test stimuli in these experiments may have only been experienced at one location, exemplars of these object categories would have been experienced at a wide variety of retinal locations through everyday experience, and this may have been necessary in order to support translation tolerance for the stimuli used in these experiments.





As illustrated in Figure 1, this led to a number of studies that assessed on-line translation tolerance for a range of unfamiliar stimuli, with highly mixed results.  In many cases these behavioural and physiological studies revealed that on-line translation tolerance was much reduced following just a few degrees of translation (Figures 1a-d).  At the same time, near complete on-line translation tolerance has been observed following shifts of 8° (Dill & Edelman, 2001; Figure 1e). Han et al. (2020) also observed near complete on-line translation tolerance when stimuli were first presented at fixation and then shifted 7° (Figure 1f), but tolerance was much more limited when the stimuli were first presented in peripheral vision and then shifted 7° to fixation (Figure 1g).  Bowers et al. (2016) demonstrated online translation tolerance at the most distal locations to date, with high performance at displacements up to 13° (Figure 1h). One notable feature of many previous studies is that they used novel stimuli that are very unlike real objects and often the stimuli differed from one another in only fine perceptual detail which may force the visual system to rely on low-level visual representations that are retinotopically constrained (e.g., Figure 1a-c).  This might be relevant to explaining the range of findings given that greater on-line translation tolerance has been observed for novel stimuli that were structurally more similar to real objects (Figure 1e) or which were designed to differ from one another in their configurational properties rather than fine details (Figure 1f-h).





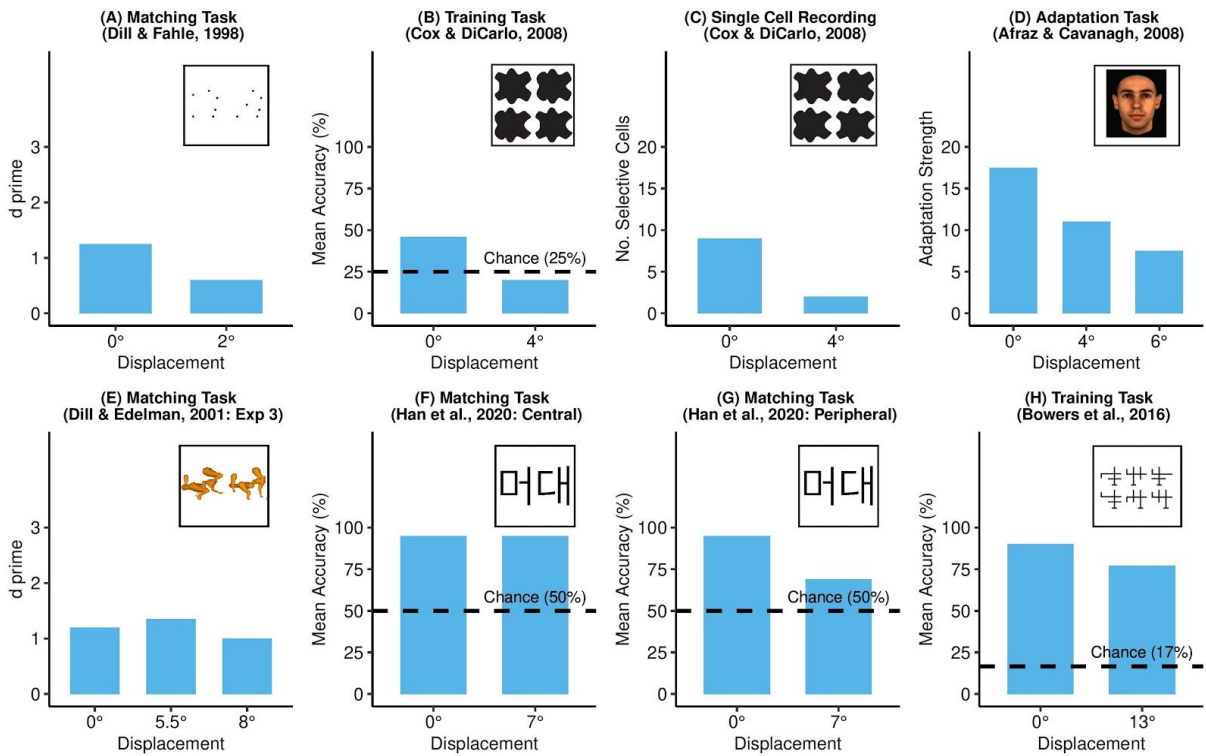

*Figure 1.* Behavioral investigations of translation tolerance, adapted from Kravitz et al. (2008). **(A)** In a 'same-different' matching task, it was significantly easier to discriminate a test-image as 'same' or 'different' to a probe image when these images were presented at matched locations compared to when the test image was displaced by just 2° (Dill & Fahle, 1998). **(B&C)** In a training task, Cox and DiCarlo (2008) demonstrated that an adult rhesus monkey performed at chance-levels (25%) when required to identify four novel objects that were displaced just 4° from the trained location, concluding that the "behavioural failure to position-generalize is caused by the monkey's reliance on a non-position tolerant visual neuronal representation" (p. 10053). In a single-cell recording study, Cox & DiCarlo (2008) also detected significantly more selective cells when novel objects were presented at the trained position compared to the displaced position. **(D)** In an adaptation task (Afraz & Cavanagh, 2008), adaptation effects (i.e., when exposure to a face alters the perception of a subsequently presented face) were inhibited after the image was displaced by 4 °, and more-so when displaced by 6°. **(E)** Using stimuli composed of scrambled animal parts, Dill & Edelman (2001) showed on-line translation tolerance over displacements of 8°. **(G&H)** Han et al. (2020) showed on-line translation tolerance over displacements of 7° although performance reduced when stimuli were trained in peripheral vision (panel H) as opposed to at fixation (panel G). **(I)** In Bowers et al. (2016), participants performed well above chance at the largest displacements to date, up to 13°.





With regard to computational modelling, most researchers have only considered trained translation tolerance. For example, a number of models of visual word identification support robust tolerance after training each word at each location (Dandurand et al., 2013; Di Bono & Zorzi, 2013). This is also the case with CNNs widely used in computer science that are often described as the best current theory of object recognition in humans (e.g., Kubilius, Kar, Schmidt, & DiCarlo, 2018). Although the design of CNNs (both the convolution and the pooling layers) are claimed to support translation invariance (for introduction see O'Shea & Nash, 2015), these models are generally trained to categorize images by training multiple exemplars of each image category at multiple spatial locations. Indeed, the standard training procedure for CNNs is to present each image across a range of positions, scales, and poses, a procedure called "data augmentation".

We are only aware of a few cases in which modellers have assessed on-line translation tolerance, and in most cases, only a limited degree of tolerance was observed. A biologically inspired neural network model called VisNet showed 100% accuracy at untrained locations for simple stimuli (Elliffe et al., 2002), but only when each stimulus was trained at multiple other spatial locations (after training in 7 locations the model generalized to an $8^{th}$ and $9^{th}$ location), and the authors only tested small translations (8 pixels in a 128x128 "retina"). This small degree of on-line tolerance was thought to provide a reasonable description of human on-line translation tolerance. The HMAX (Riesenhuber & Poggio, 1999) model showed 100% accuracy for translations up to 4° from the trained location, but its performance more than halved when objects were translated by distances equivalent to 7° and the model was not tested beyond that point. The authors also claimed that HMAX showed "the same scale and position invariance properties as the view-tuned IT neurons





described by Logothetis et al. (1995) using the same stimuli" (Riesenhuber & Poggio, 2002, p.163). There are also a few cases in which CNNs were trained on images at one retinal location and tested at another, and in most cases, highly limited on-line translation tolerance was observed (Chen et al., 2017; Furukawa, 2017; Kauderer-Abrams, 2017; Qi, 2018). However, Han et al. (2020) observed near perfect translation invariance over 7° for Korean letters in a standard CNN (see Section 3 for some more details regarding on-line tolerance in CNNs). Clearly, both the behavioral and modelling results are mixed.

Here we report a series of behavioural studies that demonstrate more extreme on-line tolerance compared to previous research and a set of simulations that examine the capacity of CNNs to support on-line translation tolerance. We show that a standard CNN (VGG16; Simonyan & Zisserman, 2014) only supports robust on-line tolerance for novel stimuli when pretrained on another set of stimuli presented at multiple retinal locations. That is, trained-tolerance for one set of stimuli led to on-line tolerance for another set of stimuli. We also show that robust on-line translation tolerance can be achieved without any pretraining by modifying the architecture of a CNN (by adding a Global Average Pooling layer to the network that generates larger receptive fields). Our findings challenge the common claim that human on-line translation tolerance is highly limited, help explain the mixed set of on-line translation tolerance results reported in CNNs, and show that standard CNNs can account for human on-line translation tolerance.





## 2.        Psychophysical Studies:  Assessing on-translation tolerance in the human visual system

        Four gaze-contingent eye-tracking studies are reported, and include the following critical design features. First, we used 24 images of naturalistic novel 3D objects organized into pairs composed of similar parts arranged in different global configurations, with one member of each pair was assigned to category A, the other to Category B (see Figure 2, Panel A; Leek, Roberts, Oliver, Cristino, & Pegna, 2016).  This should encourage participants to learn the complete objects rather than just the parts when categorizing them.  Note, previous studies have used a smaller number of novel 2D stimuli that may often have been classified on the basis of local object features (such as those depicted in Figure 1a-c).  Second, participants learned to identify the novel objects that were projected one (or two) retinal locations before being tested at novel locations.  That is, participants learned new object representations in long-term memory during training, and we assessed whether these long-term object codes supported extensive on-line tolerance.  By contrast, many previous studies did not involve learning any new representations, but rather, required matching of stimuli presented in quick succession in short-term memory (e.g., Dill & Fahle, 1997, Dill & Edelman, 2001; Han et al., 2020).  Third, we included test conditions in which objects were presented for 100ms durations, reducing the likelihood that participants adopted artificial strategies at test (e.g., slowly searching for a set of features diagnostic of category membership using covert attention). Finally, in order to test the limits of on-line translation tolerance, our experiments used displacements of up to 18° (see Experiment 4), which is larger than any previous experiment. Note that the Bowers et al. (2016) experiments reporting





robust on-line translation tolerance at 13° included a smaller number of less realistic 2D objects that were displayed for an extended time at test. Accordingly, the current studies provide a much stronger test of on-line translation tolerance.

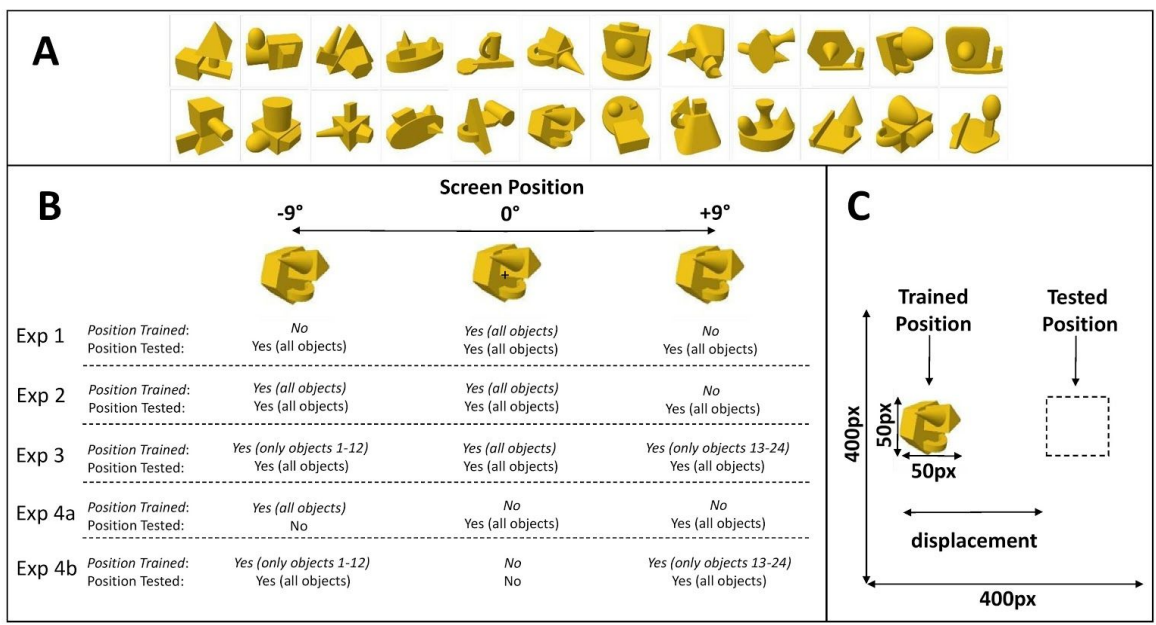

*Figure 2. (A) 24 novel objects used in behavioral studies and CNN simulations.* Each column contains a pair of objects that are matched for similar local features, but which differ in global configuration. One member of each pair was randomly assigned the label 'A' and the other was assigned 'B'. *(B) Screen positions used at training and test in behavioral experiments (fixation was always at centre).* In *Experiment 1a, 1b, & 1c*, all objects were trained at 0°, and tested at 0°, +3°, +6°, and +9° (for space reasons, +3° and +6° are not illustrated). In *Experiment 2*, all objects were trained at -9° and 0°, then tested at -9°, 0° and +9°. In *Experiment 3*, twelve objects were trained at -9° and 0°, and the other twelve were trained at 0° and +9°; all 24 objects were then tested at -9°, 0° and +9°. In *Experiments 4a* and *4b*, objects were never trained at the 0° position, and thus, novel test presentations at 9° were displaced by 18°. *(C) Illustration of procedure used to test on-line translation tolerance with CNN simulations.* A CNN (VGG16) (Simonyan & Zisserman, 2014) was trained to classify 224x224 px images containing 40x40 px novel objects; at test, the novel objects were displaced from the trained position. The precise training method used for the CNN was manipulated by crossing a number of factors, most notably: (i) Pretraining (VGG16 pretrained on ImageNet vs. VGG16 trained only to classify the 24 novel objects), (ii) Global Pooling (Global Average Pooling vs. No Global Average Pooling).





## 2.1    General Method for Psychophysical Studies (Experiments 1-4)

### 2.1.1    Ethics Statement

The study was approved by the University of Bristol Faculty of Science Ethics Committee and was in accordance with the provisions of the World Medical Association Declaration of Helsinki. All participants were recruited from the University of Bristol's course credit scheme for Psychology students.

### 2.1.2    Participants

10 participants completed each experiment (1a, 1b, 1c, 2, 3, 4a, 4b), giving 70 in total (48 female; median age = 20). Across experiments, 17 additional participants were excluded due to failure to complete the training phase within 90 minutes. The sample size was chosen based on previous psychophysical experiments that have used identification tasks to examine translation invariance (i.e., studies reported in Figure 1).

### 2.1.3    Equipment

Eye-movements were monitored using the Eyelink 1000 plus system (SR Research). Stimuli were presented using Psychopy v1.85.3 (Peirce & MacAskill, 2018) (platform: Linux-Ubuntu), and on a Viewpixx 3D Lite monitor running at 120Hz with a spatial resolution of 1920 x 1080 pixels (screen width = 53cm), at a distance of 70cm.

### 2.1.4    Procedure

In the learning phase of the experiment participants were trained to categorize the 24 objects as 'A' or 'B'. Each object was presented one-by-one and occupied 5° x 5° of visual





angle. Participants were required to maintain their gaze on a centrally located fixation-cross for 1000ms for an object to appear. If gaze moved 1.5° beyond the fixation-cross, a mask replaced the object. The learning task was split into two phases: (i) *Familiarization.* Each presentation of an object was accompanied by a sound-file indicating its category (A or B). (ii) *Training.* Each object was presented again but without the sound file and participants pressed a button to indicate each image's category. Audio feedback was then provided. The training phase continued until the participant correctly identified each object consecutively (in most experiments this required 24/24 consecutive correct answers). After the learning phase, participants completed the test phase with the 24 objects presented once in a random order in each test-block with no feedback. Again, if gaze moved 1.5° beyond the fixation-cross the object was immediately replaced with a mask. The specific details of the training and test phases in each experiment are provided in the relevant subsection below and summarised in *Table S1* (Supplementary Information). Figure 2 (panel B) provides an illustration of the training and test positions used in each experiment.

## 2.2     Experiment 1

### 2.2.1   Procedure

During the familiarization and training phases in Experiment 1a and 1b, images remained on the screen until participants responded or for just 100ms, depending on the learning block (see *Table S1*, Supplementary Information). All 24 objects were trained at the central location (at fixation) and 24/24 consecutive correct answers were required to progress to the next learning block or test phase. At test, images remained on the screen until participants responded in Experiment 1a, and for 100ms in Experiment 1b in order to reduce





possible response strategies. Experiment 1c was the same as Experiment 1b except that stimuli were presented only for 100ms in the familiarization and training phases.

### 2.2.2    Results.

Data for all experiments can be downloaded at https://osf.io/jahm9/. Table 1 shows the accuracy with which participants categorised novel objects at each test position. Performance was excellent at untrained retinal-positions (chance is 50%). Even at the most distal untrained position (9°), objects were recognised with a mean accuracy of 94% when unlimited time was afforded at test (*Experiment 1a*), and although translation tolerance was reduced when stimuli were presented for 100ms at test (*Experiments 1b, 1c*), accuracy was still at least 80% when at 9° displacement.   We carried out Bayesian Hypothesis paired sample t-tests, conducted in JASP  (JASP Team, 2019),[1] comparing performance at trained versus untrained locations.  In all conditions (across all experiments)  on-line translation tolerance was robust (all one sample t-tests produced Bayes Factors >1000), but in most cases, there was evidence for a decrease in accuracy following the most distal translations (all Bayes Factors for the 0° vs 9° comparison were >3).[2]  In the interest of space, reaction times for each condition are reported in Table S2 of the supplementary section (reaction times tell a largely similar story as the accuracy data reported above, even though our experiments were not designed as reaction time experiments and participants were not explicitly instructed to respond as quickly as possible).

---

[1] All Bayesian t-tests used the default prior option in JASP, that is, a Cauchy distribution with spread set to 0.707 (as recommended in Wagenmakers et al. 2018). Bayes Factor robustness plots were also obtained to ensure that Bayes Factors were stable across different prior specifications (accessible via JASP output files provided at https://osf.io/jahm9/).

[2] Bayes factors (BF) between 1 and 3 are considered weak or inconclusive evidence, BF between 3 and 10 are considered moderate evidence, and BF above 10 are considered strong evidence (see, Wagenmakers et al., 2018).

***Table 1. Mean (±SD) Accuracy and Bayes Factors in Experiments 1 to 4.*** The 'Training Locations' column specifies retinal positions at which stimuli were trained in each experiment (degrees of visual angle from fixation). The 'Displacement from Nearest Training Location' column shows the degrees by which test stimuli were displaced from the nearest training location, and Mean Accuracy (±*SD*) is indicated below each condition. For Experiments 2 and 3, 0° (C) and 0° (P) are shorthand to indicate whether a 0° displacement was in central (C) or peripheral (P) vision, respectively.

| Experiments | Training Locations | Displacement from Nearest Training Location | | | | Reduction in accuracy (Bayesian Paired Sample T-tests) | | |
|---|---|---|---|---|---|---|---|---|
| | | **0°** | **3°** | **6°** | **9°** | **0° vs 3°** | **0° vs 6°** | **0° vs 9°** |
| **Exp 1a** | **Always 0°** | **98%** *(±5)* | **97%** *(±5)* | **96%** *(±7)* | **94%** *(±8)* | 0.75 | 2.51 | 8.21 |
| **Exp 1b** | **Always 0°** | **97%** *(±3)* | **94%** *(±4)* | **90%** *(±6)* | **82%** *(±12)* | 12.87 | 43.7 | 92.17 |
| **Exp 1c** | **Always 0°** | **93%** *(±6)* | **92%** *(±4)* | **86%** *(±7)* | **80%** *(±8)* | 0.43 | 8.41 | 165 |

| | | **0° (P)** | **0° (C)** | **9°** | **0° (C) vs. 0° (P)** | **0° (C) vs. 9°** | **0° (P) vs. 9°** |
|---|---|---|---|---|---|---|---|
| **Exp 2** | **0° & -9°** | **93%** *(±4%)* | **95%** *(±5%)* | **85%** *(±9%)* | 2.27 | 17.56 | 9.68 |
| **Exp 3** | **0° & +/-9°** (Double Training) | **83%** *(±6%)* | **93%** *(±7%)* | **81%** *(±9%)* | 17.4 | 67.38 | 0.71 |

| | | **0°** | **9°** | **18°** | **9° vs. 18°** | **0° vs. 18°** |
|---|---|---|---|---|---|---|
| **Exp 4a** | **-9°** | Not Tested | **88%** *(±7%)* | **84%***(±7%)* | 1.63 | Not Tested |
| **Exp 4b** | **+/-9°** (Double Training) | **97%** *(±4%)* | Not Tested | **89%** *(±9%)* | Not Tested | 5.5 |

## 2.3   Experiment 2

Experiment 2 investigated two methodological factors from Experiment 1 that might explain the lower performance in periphery.  First, objects were always trained in foveal vision, and as a consequence, there was a confound with eccentricity and displacement: 0°





displacement was tested at fixation where visual acuity is high, whereas 3°, 6°, and 9° displacements were tested in peripheral vision where acuity is lower and within-object crowding by the constituent parts (Martelli et al., 2005) may further impede recognition. Second, the most distal training locations were always presented in the final test blocks, raising possible order effects.

### 2.3.1 Procedure

In Experiment 2 objects were trained 9° left of fixation as well as at fixation (see Figure 2). Three test locations were used: 9° left of fixation, centre of fixation (in fovea), and 9° right of fixation, giving three test conditions: '0° P' (0° displacement from peripheral training location), '0° C' (0° displacement from central training location), and '9°' (9° displacement from central training location, on the opposite side to the trained peripheral location). Comparing '0° P' to '9°' provides an assessment of on-line translation tolerance without the confound of eccentricity. Again, 24/24 consecutive correct answers were required at each training location to progress to the test phase. To control for possible order effects, the three test locations were randomly interleaved within each of six test-blocks.

### 2.3.2 Results

*Table 1* summarises the results of Experiment 2. Again robust on-line translation tolerance was observed in all locations, and there was still some reduction in performance between 0° C and the untrained (9°) test locations. Critically, this reduction was still present when comparing 0° P with 9° (BF=9.68), indicating that the reduction in performance to novel retinal locations cannot be attributed to the limitations of peripheral vision alone.





## 2.4     Experiment 3

In another attempt to observe more complete on-line translation tolerance we adapted a 'double-training' procedure from Xiao et al. (2008) that has been shown to overcome retinal specificity for low-level visual discrimination tasks. Xiao et al. demonstrated that participants who had been trained to discriminate contrasts at location 1 showed complete transfer of this ability to location 2 when they had also been trained to discriminate a different stimulus dimension (orientation) at location 2 (otherwise, enhanced contrast discrimination was location specific). Although it remains unclear why double training leads to position tolerance in these low-level perceptual discrimination tasks, it raised the obvious possibility that a similar training regime would lead to improved performance with our stimuli.

### 2.4.1   Procedure

In Experiment 3 we assessed identification of objects at novel test locations when those same locations were used for the training of other objects ('double-training'). As illustrated in Figure 2, all objects were trained in the fovea (until 24/24 consecutive correct answers were provided), then twelve objects were trained at one peripheral location, 9° from the central fixation-cross (until 12/12 consecutive answers were provided) and then the remaining 12 objects were trained at a contralateral peripheral location, 9° to the other side of the fixation-cross (until 12/12 consecutive answers were provided). The test phase was identical to that of Experiment 2: The key question was whether objects could be recognised as accurately at the peripheral location at which they had not been trained ('9°'), compared to the peripheral location at which they had been trained ('0° P'').





### 2.4.2  Results

The results of Experiment 3 are summarised in *Table 1*. The key finding is that accuracy at 9° (81%) was nearly equivalent to accuracy at 0° P  (83%) and BF for the paired-samples t-test was just 0.71, indicating there was no evidence for a difference between conditions even though objects were presented for just 100ms at test.  It is perhaps also worth noting that performance at 0 P° (83%) was 10% lower than the equivalent test condition in Experiment 2 (93%).  This is likely the consequence of our more lenient training criteria used at peripheral locations in double training (12/12 consecutive correct answers required at both 0° P locations as opposed to 24/24 consecutive correct answers at one 0° P location required in previous experiments - see Section *2.4.1*).

### 2.5  Experiment 4

Experiment 4 examined whether the robust on-line translation reported above could be extended to locations as distal as 18° from the trained location, which is larger than any previous demonstration of on-line translation tolerance (see Introduction). Experiment 4a investigated this question using a paradigm similar to Experiment 1 and 2 (i.e., without double training) and Experiment 4b investigated this question using double training (see Figure 2).

### 2.5.1  Procedure

In order to displace objects by 18° images were presented at one peripheral location during training, namely, 9° right or left of central fixation. Experiment 4a followed the same training procedure as Experiment 2 except that stimuli were trained at one peripheral location





only and never at fixation. Two test-blocks were used, one in which test stimuli were presented at fixation (9° displacement), and one which stimuli were tested on the opposite side of fixation (18° displacement). For Experiment 4b we used a 'double training' procedure (similar to Experiment 3), with 12 images trained 9° to the right, and the remaining 12 were trained 9° to the left of central fixation. Furthermore, in an attempt to boost performance compared to Experiment 3, 'left' and 'right' presentations were randomly interleaved within a block of 24 presentations (as opposed to being presented in separate blocks of 12) and participants were required to complete two separate loops of 24/24 consecutive correct answers (as opposed to 12/12). At test, objects were tested at two test locations: 9° left, and 9° right of fixation, giving two test conditions: '0° P' (0° displacement from peripheral trained location) and '18°' locations (18° displacement from the opposite peripheral location, i.e., 9° from central fixation) (see *Table S1, Supplementary Information*).

### 2.5.2   Results

Results of Experiment 4a and 4b are summarised in *Table 1*. In Experiment 4a participants correctly categorized 84% of objects following a displacement of 18° compared to 88% following a displacement of 9°, with a Bayesian paired-samples t-test indicating inconclusive evidence of a difference between these test locations. Note, the slightly higher performance following 9° translation may reflect that testing took place at fixation in this condition.  In Experiment 4b there was a larger drop of 8% following a displacement of 18° (89%) compared to the 0° P condition (97%).  Nevertheless, it is worth noting that 5 of 10 participants performed over 90% following an extreme displacement of 18°, with one participant scoring 96%.





**2.6     Familiarity Ratings of Objects Used in Experiments 1-4**

Although the experiments outlined above employed design constraints to minimise the role of semantics in mediating performance (i.e., rather than using familiar objects we used novel objects that had similar local features but different global configurations, and were presented for just 100ms at test - and at training in the case of Experiment 1c), it is difficult to rule this out completely. Indeed, O'Regan & Nazir (1990) note that even dot-patterns (which were not used in the current study given how unnaturalistic and difficult to differentiate they are) are susceptible to semantic strategies: "when asking the subjects afterwards what the nature of the target was, some of them gave a global description of the target as being like a chessman having something round on its head, or a bizarre telephone" (p.99).

To test the extent to which the current set of stimuli were considered novel, a new group of 20 participants were recruited via *www.prolific.ac* (Palan & Schitter, 2018), and completed an online questionnaire using *www.gorilla.sc*. Participants were instructed to "*rate the extent to which each novel object resembles a familiar object on a 5-point scale (i.e., does the novel object remind you of a particular known-object in any way?)"* where '1' indicated no resemblance to any familiar object, and '5' indicated strong resemblance to a familiar object. The mean familiarity score for the 24 novel objects was 1.91 (SD=0.34). *Figure 3* plots the familiarity score for each object against the mean accuracy score at each displaced location in the eyetracking experiments (the final panel collapses across all displaced conditions).  As is clear from these graphs there is a relation between the judged familiarity of the objects and mean accuracy, but the point we would emphasize is that robust on-line tolerance was still observed with stimuli that were given the lowest familiarity ratings.





Indeed, in the two experiments that assessed on-line translation tolerance at 18°, the correlation between familiarity and translation tolerance was non-significant (r = .13, p=0.54), and performance was approaching 90% for the 5 stimuli that were given lowest familiarity ratings. Clearly, robust on-line translation tolerance extends to objects that were judged to be completely unfamiliar.

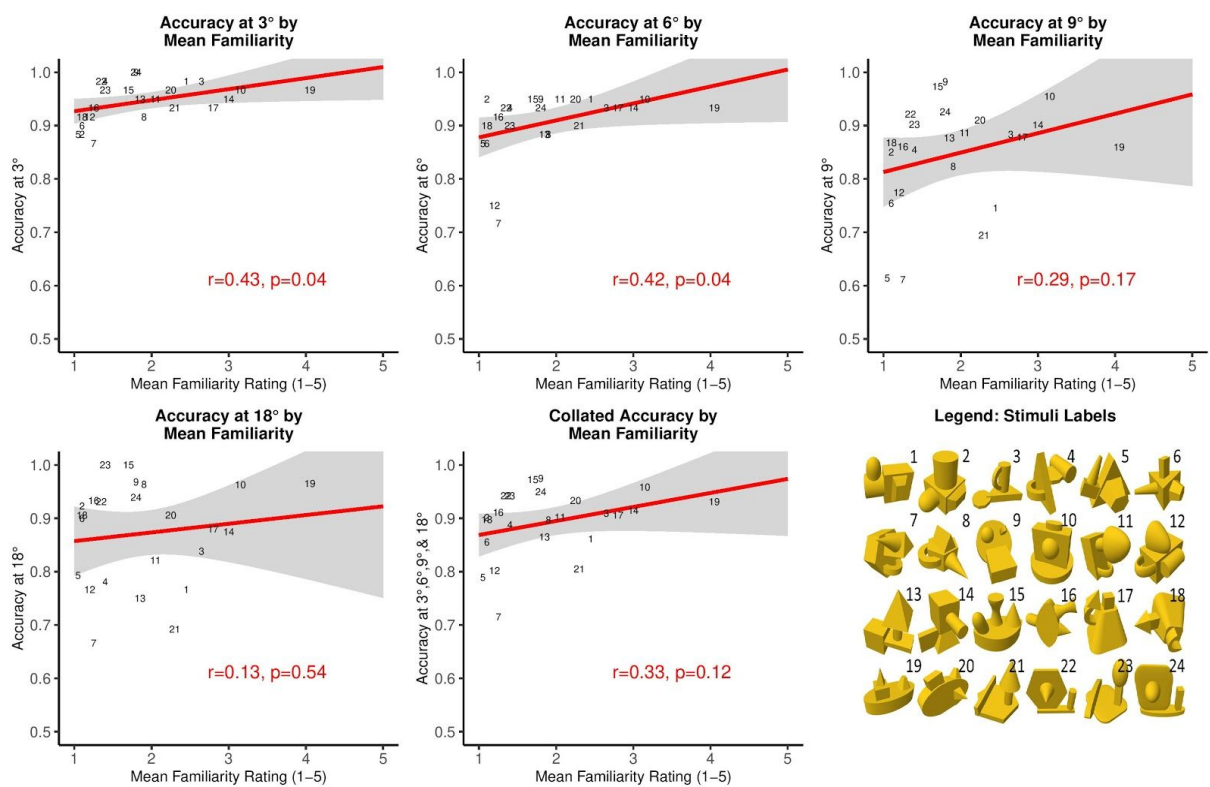

*Figure 3.* Each scatterplot shows the Mean Familiarity Score of each item (averaged over participants) on the x-axis, and the Mean Accuracy Scores of each item (averaged over participants) on the y-axis. Each plot shows Mean Accuracy Scores at a specific displacement (3°, 6°, 9°, 18°, collated) and labels each data-point so that the performance of each item can be compared across different displacements (the rank order of items is generally respected, e.g., images 7 and 12 are usually amongst the least accurate, whilst 9 and 15 are usually the most accurate). The shaded grey zone illustrates the 95% Confidence Interval for the regression line (drawn using the geom_smooth function from the ggplot2 R library) (Wickham 2016) .





### 3.  Modelling translation tolerance in CNNs

Similar to the behavioral findings, computational studies of on-line translation tolerance have been mixed, with different models supporting a range of outcomes, from near zero to near complete translation tolerance. *Table 2* details the range of outcomes with standard CNN models along with some key differences in the simulations, namely, the nature of the test stimuli, whether the models were pretrained to classify other stimuli across a range of locations, whether the test stimuli were trained in a single location or "jittered" over a small range of locations, and whether the model included a Global Average Pooling layer (see below). The column called *Largest Displacement At Test* indicates the most distal displacement from the trained location (in pixels) at which the model was tested, and the model's accuracy at that displacement is shown in the column called *Performance at Largest Displacement*. Strikingly, four out of five CNNs only displaced objects by distances that were smaller than the object's own dimensions (e.g., Chen et al. displaced 36x36 images by up to 30 pixels), and most of these showed dramatically reduced performance at that displaced location (e.g., Chen et al.'s model performed at chance-levels following 30 pixel displacements). The only exception to this was the Han et al. (2020) CNN that supported robust on-line translation tolerance for untrained Korean letters at displacements that were up to 7 times the width of the letters. Note, only the Han et al. model was trained to classify a different set of stimuli (digits from the MNIST dataset; LeCun, 1998) across multiple locations, suggesting that pretraining may be a critical factor. That is, CNNs may need to learn trained tolerance on one set of stimuli before supporting on-line tolerance for a novel set of stimuli.

In addition, it is worth noting that none of the above CNNs included a Global Average





Pooling (GAP) mechanism (Lin, Chen, & Yan, 2013) designed to provide larger receptive fields that cover the whole visual field. GAP is a hard-wired mechanism applied to each individual feature map of the final convolutional layer, and averages the values of each feature map into a single value that covers the whole visual field. Given that a GAP layer is commonly added to CNNs in order to make CNNs more robust to spatial translations of the input, this seems a relevant factor to consider as well. In the simulations below we assessed on-line translation tolerance by training the models on the same set of 3D objects at one retinal location and then testing the model at novel locations while varying three factors: a) pretraining vs. no pretraining, b) jitter vs. no-jitter on test stimuli, and c) GAP vs no-GAP.

***Table 2.*** **Previous studies that have examined translation tolerance in CNNs when restricting the location of the training image.**

| Study | Training/ Test Stimuli | Jittered Training Stimuli | Pretrained on Other Datasets | GAP | Accuracy at Trained Location | Largest Displacement at Test | Performance at Largest Displacement (chance-level was 10% unless stated) |
|---|---|---|---|---|---|---|---|
| Kauderer-Abrams (2017) | MNIST (28x28px) | No | No | No | 100% | ±10px | 10% |
| | | Yes (±10px) | No | No | 100% | ±10px | 60% |
| Qi (2017) | MNIST (28x28px) | No | No | No | 100% | ±18px | 42% |
| | | Yes (±6px) | No | No | 100% | ±18px | 98% |
| Furukawa (2017) | SAR satellite images (104x104px) | No | No | No | 100% | ±10px | 50% (chance =20%) |
| | | Yes (±8px) | No | No | 100% | ±10px | 80% (chance =20%) |
| Chen et al. (2017) | MNIST (36x36px) | No | No | No | 100% | ±30px | 10% |
| Han et al. (2020) | 24 Korean Letters (450x450px) | No | Yes (MNIST ±3150px) | No | 100% | ±3150px | 95% (chance =50%) |





## 3.1 Methods for Modelling translation tolerance in CNNs

We systematically investigated on-line tolerance in a popular CNN (VGG16; Simonyan & Zisserman, 2014) by training the network to classify the 24 'Leek' images (*Figure 2*) as 'A' or 'B' at restricted locations, and then testing its accuracy at displaced locations equivalent to the psychophysical studies. As illustrated in *Figure 2 (panel C)*, the Leek images were 40x40 pixels and were presented within 224x224 pixel space to allow for relatively large displacements at test (compared to most previous CNN investigations). In all simulations, training continued until the model reached 100% accuracy. We manipulated and crossed the following three factors (giving eight simulations in total):

*(i) No Pretraining vs Pretraining.* The VGG16 network was either (a) trained from scratch on the 24 Leek images, or (b) pre-trained on ImageNet (Russakovsky et al., 2015) (by definition, this meant experiencing exemplars of categories from ImageNet across a range of retinal locations) and then trained to classify the 24 Leek images using the existing visual representations, a procedure known as 'transfer learning'.

*(ii) No Jitter vs. Jitter.* The 24 Leek stimuli were either trained at (a) the trained location only, or (b) a limited range of locations (Jitter Condition), randomly displaced from the trained location by up to 20 pixels each side (thus introducing some variability as well as mimicking the behavioral studies in which fixations were free to vary by 1.5 degrees before image was masked).





*(iii) Global Average Pooling (GAP) vs. No GAP.* The VGG16 network either (a) uses GAP on the resulting feature map from the final convolutional layer of VGG16, or (b) does not use GAP (akin to the previous examinations of on-line translation tolerance in CNNs; see *Table 2*). At test, the Leek stimuli were displaced up to 160 pixels from the trained location. Therefore, the highest displacement was 4 times the width of the 40x40 pixel stimuli, similar to our psychophysical studies (which displaced 5°x5° images by up to 3.6 times their width). The model's accuracy for each condition was averaged over 20 replications.

### 3.1.1   Results

*Figure 4* summarises the outcome of simulations when only horizontal displacements were used, consistent with our psychophysical experiments. When the CNN model included a GAP layer perfect online translation invariance was observed across all displacements, regardless of the pretraining or jitter (dashed red line in all four panels of Figure 4B). By contrast, when a standard CNN model was used (solid lines in Figure 4B), robust translation tolerance was only observed when the model was pretrained on ImageNet (Figure 4B, bottom panels), and following pretraining, there was a small reduction in performance following translations, with performance dropping from ~100% to ~85%. Jitter only had a minor effect on the pretrained model, but improved performance on the untrained model in the region of the jitter. To facilitate comparison between human and CNN performance, the black dashed line depicted in the bottom right panel of Figure 4b plots mean human performance (collapsing over experiments) after converting each displacement (0°, 3°, 6°, 9°, 18°) to an





equivalent measurement in pixels based on the proportion of the image size.[3] Clearly the

pretrained standard CNNs (without GAP) accounts for the human data most closely, and is

consistent with the fact that the humans in our experiment had extensive previous experience

seeing familiar objects at multiple retinal locations.

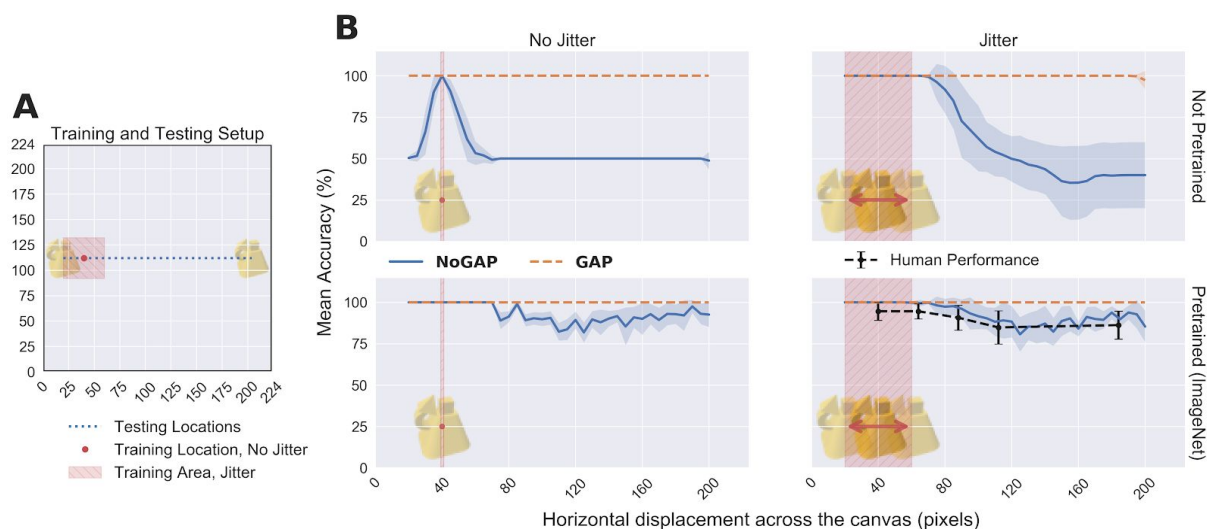

*Figure 4.* **Mean Accuracy of CNN when classifying Leek (2016) stimuli over large translations. (A)**

Illustration of the experimental setup, showing the trained and tested locations. The red dot represents the

location where the training stimuli were centered for the 'No Jitter' condition and the red shaded area represents

the locations at which the training stimuli were centered for the 'Jitter' conditions. At test, stimuli were

displaced by up to 160 pixels, which is four times the width of the 40x40 pixel stimuli, thus corresponding

roughly to the displacements used in our psychophysical experiments. **(B)** Illustration of CNN accuracy at test

locations, crossing jitter, pretraining and GAP. Shaded areas represent one standard deviation. Black dashed line

in the bottom right panel depicts human performance at analogous test locations.

---

[3] For example, in psychophysical experiments, all objects were 5° wide so displacements of
3°,6°,9°,18° were, respectively, x0.6, x1.2, x1.8, x3.6 the size of the image and thus, the
equivalent displacements of the 40px image used in simulations are 24px, 48px,72px,144px.





We also repeated the simulations across a greater range of displacements. As illustrated in *Figure 5*, each Leek image was tested on a 19x19 grid in the canvas, centering every stimulus at each point of the grid. The results were averaged across 20 replications, and the untested points in the canvas were estimated through cubic interpolation. The results highlight even more clearly the limited on-line translation tolerance obtained with standard DCNNs without pretraining, the extreme on-line translation tolerance obtained with standard DCNNs that were pretrained on ImageNet, the limited impact of jitter, and the compete online translation invariance obtained with DCNNs that include a GAP layer regardless of pretraining.

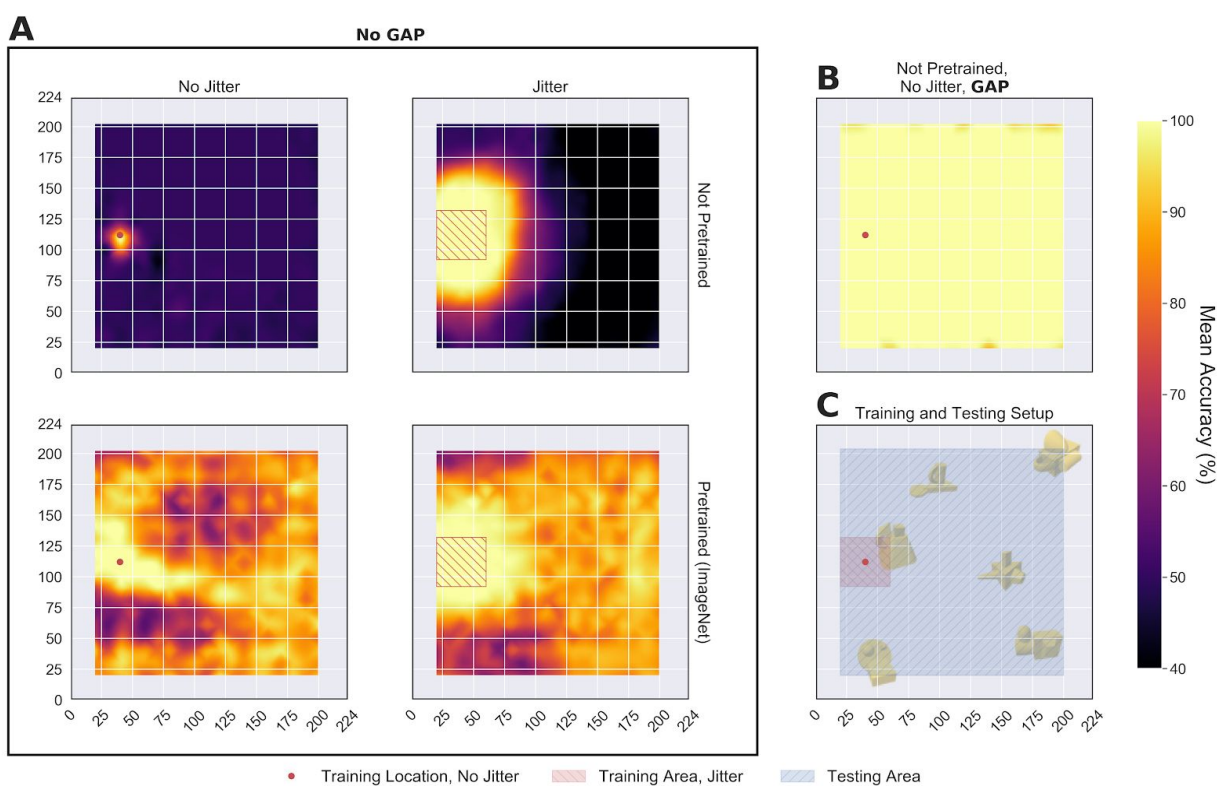

*Figure 5.* **Mean CNN accuracy across 20 runs on the whole canvas, for GAP and no GAP conditions.** Test results are expanded across the whole canvas. **(A)** No GAP conditions, with/without jitter and with/without





pretraining on ImageNet. When no pretraining was performed, the network does not generalize on translation much further than the trained locations (upper panels). When the network was pretrained, the network could generalize much better across the whole canvas (bottom panels). In these panels, the small areas with lower accuracy might be the results of random features of the pretraining dataset (e.g. photographer bias). **(B)** With GAP, the accuracy was at ceiling everywhere on the canvas. Results for the conditions with pretraning and with jitter presented a similarly high accuracy, and the plots are therefore omitted. **(C)** Diagram of the experimental setup when tested across the whole canvas. The shaded areas represent the locations of the *center* of the image.

## 4.    Discussion

In a series of behavioral experiments we demonstrate that participants trained to recognise images of novel 3D objects at one retinal position can recognise the same objects at untrained distal retinal-locations with high accuracy (up to 18°).  These findings challenge the common claim that on-line translation tolerance is highly limited (Chen et al., 2017; Cox, & DiCarlo, 2008; for review, see Kravitz et al., 2008), but are consistent with early priming studies (e.g., Biederman & Cooper, 1991), as well as a number of more recent studies that observed robust translation tolerance up to 13°  (e.g., Bowers et al., 2016).  Similarly, in a series of simulation studies, we have identified conditions in which CNNs can support this degree of on-line translation tolerance, namely, when the model was pretrained on another set of stimuli presented at multiple locations, or when they included a hard-wired ("innate") mechanism that forces the networks to learn large receptive fields, in this case, a GAP layer (although other mechanisms may also support these large receptive fields).  The results help reconcile the mixed findings reported in the literature (as summarized in Table 2), and highlight how pooling and convolutional operations employed in CNNs can account for the robust on-line translation tolerance we observed.





It should be noted that we consistently observed a small decrease in performance across behavioral experiments when novel objects were presented to novel locations at test. This was the case even when we controlled for the eccentricity of trained and novel locations (Experiment 2), and when we employed a double training procedure so that participants were practiced at identifying stimuli at the critical test locations (Experiments 3 and 4b). For this reason we can only conclude that the visual system supports extreme on-line translation *tolerance* rather than complete on-line translation *invariance*. Importantly, we consistently obtained robust translation tolerance when test stimuli were presented for 100 ms, suggesting that bottom-up perceptual processes played an important role in this tolerance. This is consistent with our simulation studies that all used CNNs that operate in a bottom-up manner.

With regards to our simulation studies, we found that a standard CNN (without GAP) needs to be pretrained to classify another set of stimuli (in this case images from the ImageNet dataset) presented at multiple retinal locations in order to manifest robust on-line tolerance for our novel images of 3D test stimuli. That is, the standard CNNs only exhibited extensive on-line tolerance after acquiring trained-tolerance for a different set of stimuli. It is of course the case that the participants in our behavioral experiments were exposed to many familiar images at multiple retinal locations prior to learning the novel 3D images, so this constraint on on-line tolerance is psychologically plausible. Furthermore, the pretrained CNNs seem to provide a reasonable account of human performance, showing near-perfect accuracy at nearby displacements, and ~10% reduction in accuracy at the more distal locations. By contrast, the models with a GAP layer did not need to be pretrained in order to support on-line translation invariance. Although the CNN with a GAP layer appears to





provide a better solution to on-line translation tolerance from a machine learning perspective, it does not capture the limitations of human performance.

As far as we are aware, no one has documented this link between trained- and on-line tolerance in models of vision. Indeed, as noted earlier, most models and theories of word and object identification reject robust on-line tolerance and instead assume that trained tolerance explains how humans (and monkeys) identify images across a wide range of retinal locations (e.g., Chen et al., 2017; Cox & DiCarlo, 2008; Dandurand et al., 2013; Di Bono & Zorzi, 2013; Edelman & Intrator, 2003; Elliffe et al., 2002). Although our findings challenge this conclusion, our findings are consistent with the more general point that trained-tolerance on one set of stimuli (i.e., pretraining on other categories) is a prerequisite for the on-line tolerance that we observed. Why pretraining is required for standard DCNNs but not with DCNNs with a GAP layer is an interesting question for future research. One possibility is that the pretraining allows the model to discover invariant low-level features which it then uses to categorize novel stimuli.

In other work, Han et al. (2020) observed robust on-line translation tolerance for a standard pretrained CNN. However, they argued that it provided a poor account of their behavioral findings. These authors observed robust tolerance when novel stimuli were first presented at fixation and translated 7° to the periphery, but tolerance was much reduced when the novel stimuli were first presented in periphery (see Figure 1f & g). The standard, pretrained CNN did not capture this asymmetry. In order to explain their findings they employed an eccentricity-dependent Neural Network (or ENN), a CNN model that included multiple parallel channels that sampled the inputs at different spatial resolutions, with only the low spatial resolution channel processing images at the larger eccentricities. Although





they were able to explain the asymmetry in on-line translation tolerance with this modified

CNN architecture, we obtained no evidence of this asymmetry in our psychophysical studies,

with performance equally impressive when trained in periphery only (see Experiments 4a and

4b) as when trained at fixation only (see Experiments 1b, 1c). Why Han et al. observed a

different pattern of results is  unclear given the many methodological differences between our

behavioral studies.  Whatever the reason, their model is inconsistent with the current and past

results that report near complete on-line translation tolerance for stimuli first presented

beyond 7° eccentricity (Bowers et al., 2016; Dill & Edelman, 2001).

     In future research it will be important to explain the mixed behavioral on-line

translation tolerance results obtained across studies. For example, one possible explanation

for the mixed behavioral findings is that (at least) two representations can be used for object

recognition: a representation of shape that is invariant to position and an episodic

representation that is bound to the original viewing experience (e.g., Biederman & Cooper,

1991; Biederman et al., 2009). More research is needed to develop this theory, as well as to

investigate factors that might mediate the mixed findings. For example, the role of task and

stimulus complexity in on-line translation tolerance is yet to be systematically investigated,

although there is evidence from other domains of invariance (e.g., pose invariance) that these

factors may play a role  (e.g., Tjan & Legge 1998). Additionally, further work needs to assess

other forms of invariance in CNNs, including scale, rotation in the picture plane, rotation in

depth, etc., and compare to human performance in order to explore the similarities of human

vision and CNNs more thoroughly.  It is also worth noting that the CNN models used here

and in other similar works do not generate reaction times (RTs). In fact, even though

modelling RTs through artificial networks has been substantially explored (Bogacz et al.,





2006), there have not been many attempts to produce reaction times out of CNNs: the only example to our knowledge is Holmet et al. (2020), which used CNNs to parameterize the drift in a classic drift diffusion model. As we observed in Experiments 1a-c, the time of presentation of the stimuli might play a role in human performance, and thus an improved model would take this into account.

Overall, the current simulation studies capture the on-line translation tolerance demonstrated in our psychophysical studies, and also account for the mixed results previously reported with CNNs. Our findings of extreme on-line translation tolerance in humans and CNNs undermine models and theories that posit more limited translation tolerance.

**Supplemental Material**

In the behavioral studies, the training phase was divided into several blocks to facilitate

learning. In the first four blocks, each block used just 6 objects (3 'A' and 3 'B'). After

completing block 4, participants completed a fifth block (5a) in which all 24 objects were

presented. In some experiments, two additional training phases were used - Block 5b and

Block 5c - which were identical to 5a except for their shorter presentation times or positions.

A summary of presentation times and positions for each training and test block are provided

in Table S1, below.

*Table S1.* **Summary of Designs used in Experiments 1 to 4.**

|  | Training positions (Block) | Training Presentation Times (Block) | Displacement of Test Positions | Test Presentation Time |
|---|---|---|---|---|
| **Exp 1a** (N=10) | 0° (1-5c) | Unlimited (1-5a), 500ms (5b), 100ms (5c) | 0°,3°,3°,6°,6°,9°,9° (blocked trials of 24 x 7) | Unlimited Time |
| **Exp 1b** (N=10) | 0° (1-5c) | Unlimited (1-5a), 500ms (5b), 100ms (5c) | 0°,3°,3°,6°,6°,9°,9° (blocked trials of 24 x 7) | 100ms |
| **Exp 1c** (N=10) | 0°(1-5a) | 100ms (1-5a) | 0°,3°,3°,6°,6°,9°,9° (blocked trials of 24 x 7) | 100ms |
| **Exp 2** (N=10) | 0°(1-5b) & 9° Peripheral (5c) | Unlimited (1-5a), 100ms (5b, 5c) | 0°/0°/9° (interleaved trials of 24 x 6) | 100ms |
| **Exp 3** (N=10) | 0° (1-5b) & Both 9° Peripheral (5c) | Unlimited (1-5a), 100ms (5b, 5c) | 0°/0°/9° (interleaved trials of 24 x 6) | 100ms |
| **Exp 4a** (N=10) | 9° Peripheral (1-5c) | Unlimited (1-5a), 500ms (5b), 100ms (5c) | 9°, 18° (blocked trials of 24) | 100ms |
| **Exp 4b** (N=10) | Both 9° Peripheral (1-5c) | Unlimited (1-5a), 100ms (5b, 5c) | 0°/18° (interleaved trials of 24 x 4) | 100ms |

*Table S1.* **Training:** Training position and presentation times are specified in the first two columns, with the relevant training blocks in parentheses. For example, *Exp. 1a & 1b* present images at a central (0°) position and for unlimited time in blocks 1-5a, 500ms in block 5b, and 100ms in block 5c. *Exp. 2* presented images at 0° for unlimited time in blocks 1-5a and for 100ms in block 5b, but block 5c presented objects at one peripheral side (9° horizontal eccentricity), for 100ms. *Exp. 3* used the same training procedure as *Exp. 2* except that block 5c used a double training procedure (i.e., *both* peripheral positions were trained such that 12 objects were trained on the left side and the remaining 12 on the right side; see Experiment 3). **Test:** Test positions (displacement from nearest training position) and presentation times are specified in the final two columns. Depending on the experiment, test positions were 'blocked' (where the same position was used for all 24 trials within a block and each block used a different position) or interleaved (where one of the possible positions was randomly assigned for each trial). 'Blocked' test positions with the same displacement (e.g., *Exp. 1* uses 3° displacements for





two blocks) differed in terms of presentation side (left or right). Note that some experiments that used 'interleaved' test positions used two 0° displacements: one for a 'peripherally-trained' position, and another for a 'centrally-trained' position (see, e.g., Experiment 3).

The mean reaction times for each condition are reported in Table S2, which corresponds to the analysis of accuracy data in Table 1 of the main manuscript.

***Table S2. Mean Reaction Times (ms) (excluding responses that were >3000ms and/or incorrect) and Bayes Factors in Experiments 1 to 4.***

| Experiments | Training Locations | Displacement from Nearest Training Location | | | | Reduction in accuracy (Bayesian Paired Sample T-tests) | | |
|---|---|---|---|---|---|---|---|---|
| | | 0° | 3° | 6° | 9° | 0° vs 3° | 0° vs 6° | 0° vs 9° |
| Exp 1a | Always 0° | 1063 (±191) | 1128 (±219) | 1163 (±286) | 1211 (±260) | 2.63 | 2.15 | 47.07 |
| Exp 1b | Always 0° | 1048 (±190) | 1078 (±94) | 1109 (±141) | 1087 (±160) | 0.47 | 1.07 | 0.58 |
| Exp 1c | Always 0° | 1186 (±235) | 1286 (±232) | 1143 (±132) | 1172 (±192) | 10.5 | 0.17 | 0.27 |

| | | 0° (P) | 0° (C) | 9° | 0° (C) vs. 0° (P) | 0° (C) vs. 9° | 0° (P) vs. 9° |
|---|---|---|---|---|---|---|---|
| Exp 2 | 0° & -9° | 1031 (±183) | 1177 (±221) | 1271 (±266) | 39.5 | 220 | 14 |
| Exp 3 | 0° & +/-9° (Double Training) | 1097 (±219) | 1234 (±258) | 1329 (±272) | 19.76 | 69.49 | 9.76 |

| | | 0° | 9° | 18° | 9° vs. 18° | 0° vs. 18° |
|---|---|---|---|---|---|---|
| Exp 4a | -9° | Not Tested | 1355 (±190) | 1296 (±182) | 0.15 | Not Tested |
| Exp 4b | +/-9° (Double Training) | 1135 (±234) | Not Tested | 1350 (±236) | Not Tested | 670 |

**Table S2.** The 'Training Locations' column specifies retinal positions at which stimuli were trained in each experiment (degrees of visual angle from fixation). The 'Displacement from Nearest Training Location' column shows the degrees by which test stimuli were displaced from the nearest training location, and Mean Accuracy (±*SD*) is indicated below each condition. For Experiments 2 and 3, 0° (C) and 0° (P) are shorthand to indicate whether a 0° displacement was in central (C) or peripheral (P) vision, respectively.